\documentclass[pre,aps,amssymb,amsmath,superscriptaddress,twocolumn]{revtex4}
\usepackage{amsmath}
\usepackage{amssymb}
\usepackage{epsfig}
\usepackage{amscd}
\usepackage{graphicx,psfrag,xspace}
\usepackage{float}
\usepackage[normalem]{ulem}

\begin{document}

\title{Collision statistics for random flights with anisotropic scattering and absorption}
\author{A. Zoia}
\email{andrea.zoia@cea.fr}
\affiliation{CEA/Saclay, DEN/DANS/DM2S/SERMA/LTSD, 91191 Gif-sur-Yvette, France}
%\affiliation{Commissariat \`a l'Energie Atomique et aux Energies Alternatives, Direction de l'Energie Nucl\'eaire, D\'epartement de Mod\'elisation des Syst\`emes et Structures, Service d'Etudes des R\'eacteurs et de Math\'ematiques Appliqu\'ees, CEA/Saclay, 91191 Gif-sur-Yvette, France}
\author{E. Dumonteil}
\affiliation{CEA/Saclay, DEN/DANS/DM2S/SERMA/LTSD, 91191 Gif-sur-Yvette, France}
\author{A. Mazzolo}
\affiliation{CEA/Saclay, DEN/DANS/DM2S/SERMA/LTSD, 91191 Gif-sur-Yvette, France}

\begin{abstract}
For a broad class of random walks with anisotropic scattering kernel and absorption, we derive explicit formulas that allow expressing the moments of the collision number $n_V$ performed in a volume $V$ as a function of the particle equilibrium distribution. Our results apply to arbitrary domains $V$ and boundary conditions, and allow assessing the hitting statistics for systems where the typical displacements are comparable to the domain size, so that the diffusion limit is possibly not attained. An example is discussed for one-dimensional ($1d$) random flights with exponential displacements, where analytical calculations can be carried out.
\end{abstract}
\maketitle

\section{Introduction}

The dynamics of complex physical systems is often described in terms of `particles' undergoing random displacements, resulting either from the intrinsic stochastic nature of the underlying process, or from uncertainty~\cite{hughes, weiss}. Widespread examples arise in radiation transport, research strategies, biology and percolation through porous media, only to name a few~\cite{schlesinger, bouchaud_desorder, zoia, lecaer}. In this context, quantifying the residence time that the walkers spend inside a given domain $V$ is a key issue that has motivated a considerable research effort~\cite{redner, avraham, klafter, condamin, condamin_benichou, grebenkov, benichou_grebenkov, grebenkov_jsp, barkai, majumdar_fptd, agmon_original}.

When the domain size is large as compared to the average displacements, the walker dynamics is usually modelled by either regular Brownian motion for homogenous media, or anomalous diffusion for heterogeneous, scale-invariant media~\cite{klafter, avraham, hughes, weiss}. For Markovian transport processes, a systematic approach to assessing the residence time distribution exists, via the so-called Feynman-Kac formalism~\cite{kac_original, kac_berkeley, kac_darling, kac, majumdar_review}. Yet, full knowledge of the residence time distribution is an awkward task, and is achievable only in a limited number of cases~\cite{kac, grebenkov, berezhkovskii}, so that one has often to be content with the first few moments of the residence time~\cite{condamin_benichou, redner, agmon, agmon_lett}. A further difficulty arises when the walker typically undergoes a limited number of collisions before leaving the explored domain, and the diffusion limit is possibly not attained. This is often the case in gas dynamics, neutronics and radiative transfer, electronics, and biology~\cite{cercignani, wigner, jacoboni_book, lecaer, benichou_epl}. In all such systems, the stochastic path can be thought of as a series of straight-line flights, separated by random collisions, and a natural variable for describing the walker evolution is therefore the number of collisions $n_{V}$ within the observed volume. Application of the diffusion approximation to the counting statistics, which amounts to assuming a large number of collisions in $V$, might lead to inaccurate results~\cite{blanco}.

In a previous work, we have addressed the issue of characterizing the moments $\langle n^m_{V} \rangle$ for arbitrary geometries and boundary conditions, subject to the condition that both the source and scattering are isotropic~\cite{zdm_prl}. Here we extend those results by relaxing the isotropy hypothesis. We also distinguish the case where events are counted before or after each collision. We derive explicit formulas for the moments $\langle n^m_{V} \rangle$ by building on survival probabilities, and relate the collision statistics to the particle equilibrium distribution. Knowledge of higher order moments allows estimating the uncertainty on the average, as well as reconstructing the full distribution of the collision number. We exemplify our findings by examining a case of one-dimensional ($1d$) transport, the so called {\em rod model} with exponentially distributed displacements, where analytical calculations can be carried out.

This paper is structured as follows: in Sec.~\ref{random_flights} we briefly recall the basic properties of random flights performing anisotropic scattering and absorption. In Sec.~\ref{direct_section} we derive the moments $\langle n^m_{V} \rangle$ by a direct contruction based on survival probabilities. Then, in Sec.~\ref{diff_lim} we discuss the asymptotic results that are recovered in the large $n_V$ limit. Finally, an example is worked out in Sec.~\ref{rod_sec}, and conclusions are drawn in Sec.~\ref{conclusions}.

\section{Random flights: transport kernels and equilibrium distributions}
\label{random_flights}

Consider the random walk of a particle emitted at velocity ${\mathbf v}_0$ from a point-source ${\cal S}$ located at ${\mathbf r}_0$. At each collision, the particle can be either scattered, with probability $p({\mathbf r},{\mathbf v})$, or absorbed (in which case the trajectory terminates).

\begin{figure}[t]
   \centerline{ \epsfclipon \epsfxsize=9.0cm
\epsfbox{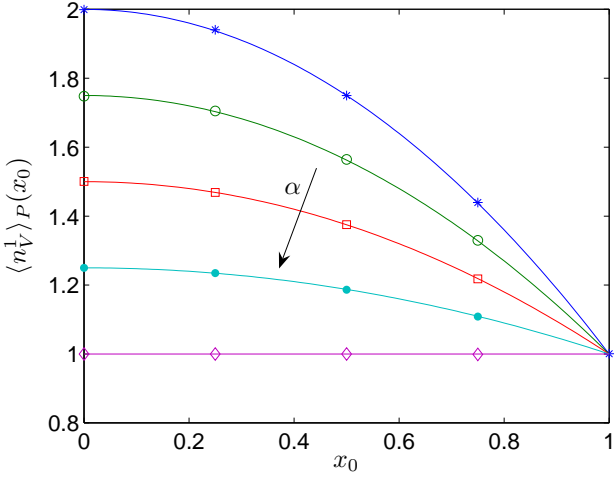} }
   \caption{(Color online) Average collision number $\langle n^1_V\rangle_{{P}}(x_0)$ for a sphere with radius $R=1$ and leakage boundary conditions. The source is isotropic, and $p=1$. Solid lines: Eq.~\eqref{moment_one_aniso}. Symbols: Monte Carlo simulations. The persistence parameter is $\alpha=0$ (dark blue stars), $\alpha=0.25$ (green circles), $\alpha=0.5$ (red squares), $\alpha=0.75$ (light blue dots), and $\alpha=1$ (violet diamonds).}
   \label{fig1}
\end{figure}

We introduce the quantity $T({\mathbf r}|{\mathbf r'},{\mathbf v'})$, namely, the probability density of performing a displacement from ${\mathbf r'}$ to ${\mathbf r}$ (in the direction of ${\mathbf v'}$), between any two collisions. Then, $\int_V T({\mathbf r}|{\mathbf r'},{\mathbf v'}) d{\mathbf r}$ represents the average number of next collisions in a volume $V$ per particle emitted at ${\mathbf r'}$, with velocity ${\mathbf v'}$~\cite{spanier, lux}. Analogously, we introduce $C({\mathbf v}|{\mathbf v'},{\mathbf r})$, namely, the conditional probability density of changing velocity from ${\mathbf v'}$ to ${\mathbf v}$, given a scattering event at ${\mathbf r}$. Usually, we can factorize $C({\mathbf v}|{\mathbf v'},{\mathbf r})=p({\mathbf r},{\mathbf v'})c({\mathbf v}|{\mathbf v'},{\mathbf r})$, where $c({\mathbf v}|{\mathbf v'},{\mathbf r})$ denotes the normalized scattering kernel. Then, $\int_V C({\mathbf v}|{\mathbf v'},{\mathbf r'}) d{\mathbf v}$ represents the average number of particles leaving a collision per incident particle entering the collision at ${\mathbf r'}$, with velocity ${\mathbf v'}$~\cite{spanier, lux}. It follows that
\begin{equation}
K({\mathbf r},{\mathbf v}|{\mathbf r'},{\mathbf v'})=T({\mathbf r}|{\mathbf r'},{\mathbf v})C({\mathbf v}|{\mathbf v'},{\mathbf r'})
\end{equation}
represents the density of particles entering the $(n+1)$-th collision with coordinates ${\mathbf r},{\mathbf v}$, having entered the $n$-th with coordinates ${\mathbf r'},{\mathbf v'}$. Inversing the order of the displacement and collision kernels, the quantity
\begin{equation}
L({\mathbf r},{\mathbf v}|{\mathbf r'},{\mathbf v'})=C({\mathbf v}|{\mathbf v'},{\mathbf r})T({\mathbf r}|{\mathbf r'},{\mathbf v'})
\end{equation}
represents the density of particles leaving the $n$-th collision with coordinates ${\mathbf r},{\mathbf v}$, having left the $(n-1)$-th with coordinates ${\mathbf r'},{\mathbf v'}$.

We can now define the displacement operator
\begin{equation}
{\mathop{\rm T}\nolimits}[ f] ({\mathbf r},{\mathbf v'})= \int T({\mathbf r}|{\mathbf r'},{\mathbf v'}) f({\mathbf r'},{\mathbf v'}) d{\mathbf r'},
\label{pi_operator}
\end{equation}
and the collision operator
\begin{equation}
{\mathop{\rm C}\nolimits}[ f] ({\mathbf r},{\mathbf v})= \int C({\mathbf v}|{\mathbf v'},{\mathbf r}) f({\mathbf r},{\mathbf v'}) d{\mathbf v'}
\label{coll_operator}
\end{equation}
for any sufficiently well-behaved $f$. Furthermore, we introduce the transport operators
\begin{equation}
{\mathop{\rm K}\nolimits}[ f] ({\mathbf r},{\mathbf v})= \int d{\mathbf r'}  \int d{\mathbf v'} K({\mathbf r},{\mathbf v}|{\mathbf r'},{\mathbf v'}) f({\mathbf r'},{\mathbf v'}),
\label{transport_operators_K}
\end{equation}
and
\begin{equation}
{\mathop{\rm L}\nolimits}[ f] ({\mathbf r},{\mathbf v})= \int d{\mathbf v'}  \int d{\mathbf r'} L({\mathbf r},{\mathbf v}|{\mathbf r'},{\mathbf v'}) f({\mathbf r'},{\mathbf v'}).
\label{transport_operator_L}
\end{equation}

\begin{figure}[t]
   \centerline{ \epsfclipon \epsfxsize=9.0cm
\epsfbox{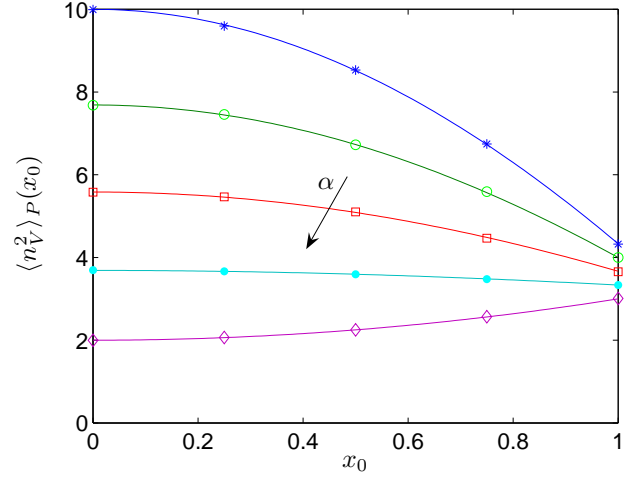} }
   \caption{(Color online) Second moment of the collision number $\langle n^2_V\rangle_{{P}}(x_0)$ for a sphere with radius $R=1$ and leakage boundary conditions. The source is isotropic, and $p=1$. Solid lines: Eq.~\eqref{moment_two_aniso}. Symbols: Monte Carlo simulations. The persistence parameter is $\alpha=0$ (dark blue stars), $\alpha=0.25$ (green circles), $\alpha=0.5$ (red squares), $\alpha=0.75$ (light blue dots), and $\alpha=1$ (violet diamonds).}
   \label{fig2}
\end{figure}

To simplify notation, we denote in the following ${\mathbf z}=\left\lbrace {\mathbf r},{\mathbf v} \right\rbrace $ the coordinates of the walker in the phase space. We introduce then the incident propagator $\Psi({\mathbf z},n|{\mathbf z}_0)$, i.e., the probability density of finding a particle entering the $n$-th collision with coordinates ${\mathbf z}$, starting from ${\mathbf z}_0$, and the outgoing propagator $\chi({\mathbf z},n|{\mathbf z}_0)$, i.e., the probability density of finding a particle exiting the $n$-th collision with coordinates ${\mathbf z}$, starting from ${\mathbf z}_0$. These quantities are related by
\begin{equation}
\chi({\mathbf z},n|{\mathbf z}_0)={\mathop{\rm C}\nolimits} \Psi({\mathbf z},n|{\mathbf z}_0)
\end{equation}
and
\begin{equation}
\Psi({\mathbf z},n|{\mathbf z}_0)={\mathop{\rm T}\nolimits}\chi({\mathbf z},n-1|{\mathbf z}_0),
\end{equation}
with $n\ge 1$, and $\chi({\mathbf z},0|{\mathbf z}_0)={\cal S}$. It follows also
\begin{equation}
\Psi({\mathbf z},n+1|{\mathbf z}_0)={\mathop{\rm K}\nolimits}\Psi({\mathbf z},n|{\mathbf z}_0)
\label{eq_psi_chi_one}
\end{equation}
and
\begin{equation}
\chi({\mathbf z},n|{\mathbf z}_0)={\mathop{\rm L}\nolimits}\chi({\mathbf z},n-1|{\mathbf z}_0).
\label{eq_psi_chi_two}
\end{equation}
In other words, knowledge of the system state ${\mathbf z}$ at $n$ is sufficient to determine the state at ${n+1}$. From Eqs.~\eqref{eq_psi_chi_one} and~\eqref{eq_psi_chi_two}, by recursion we have
\begin{equation}
\Psi({\mathbf z},n|{\mathbf z}_0)={\mathop{\rm K}\nolimits}^{n-1}{\mathop{\rm T}\nolimits}[{\cal S}]
\label{psi_n}
\end{equation}
and
\begin{equation}
\chi({\mathbf z},n|{\mathbf z}_0)={\mathop{\rm L}\nolimits}^{n}[{\cal S}],
\label{chi_n}
\end{equation}
where ${\mathop{\rm A}\nolimits}^n$ is a $n$-fold iterated operator. These relations show that the particle dynamics is entirely defined in terms of the two kernels $C$ and $T$.

\begin{figure}[t]
   \centerline{ \epsfclipon \epsfxsize=9.0cm
\epsfbox{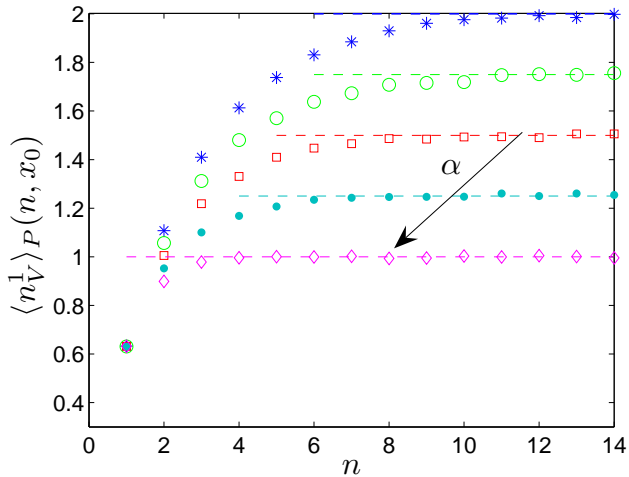} }
   \caption{(Color online) Average collision number $\langle n^1_V\rangle_{{P}}(n,x_0)$ for a sphere with radius $R=1$ and leakage boundary conditions. The source is isotropic ($x_0=0$), and $p=1$. Dashed lines: Eq.~\eqref{moment_one_aniso}. Symbols: Monte Carlo simulations. The persistence parameter is $\alpha=0$ (dark blue stars), $\alpha=0.25$ (green circles), $\alpha=0.5$ (red squares), $\alpha=0.75$ (light blue dots), and $\alpha=1$ (violet diamonds).}
   \label{fig3}
\end{figure}

We introduce now the incident and outgoing collision densities, respectively,
\begin{eqnarray}
\Psi({\mathbf z}|{\mathbf z}_0)=\lim_{N \to \infty} \sum_{n=1}^N \Psi({\mathbf z},n|{\mathbf z}_0),\nonumber \\
\chi({\mathbf z}|{\mathbf z}_0)=\lim_{N \to \infty} \sum_{n=0}^N \chi({\mathbf z},n|{\mathbf z}_0),
\end{eqnarray}
which can be interpreted as the particle stationary distributions~\cite{spanier, lux}. We can associate to the collision densities their respective operators, namely,
\begin{eqnarray}
{\mathop{\rm \Psi}\nolimits}[f]({\mathbf z}) = \int \Psi({\mathbf z}|{\mathbf z'}) f({\mathbf z'}) d{\mathbf z'},\nonumber \\
{\mathop{\rm \chi}\nolimits}[f]({\mathbf z}) = \int \chi({\mathbf z}|{\mathbf z'}) f({\mathbf z'}) d{\mathbf z'}.
\end{eqnarray}
In particular, ${\mathop{\rm \Psi}\nolimits}[{\cal S}]=\Psi({\mathbf z}|{\mathbf z}_0)$, and ${\mathop{\rm \chi}\nolimits}[{\cal S}]=\chi({\mathbf z}|{\mathbf z}_0)$. Now, by making use of the formal Neumann series (see Eq.~\eqref{neumann_def}), from Eqs.~\eqref{psi_n} and~\eqref{chi_n} we have then
\begin{equation}
{\mathop{\rm \Psi}\nolimits}[f]({\mathbf z})= \frac{{\mathop{\rm T}\nolimits}}{{\mathop{\rm I}\nolimits}-{\mathop{\rm K}\nolimits}}[f] ({\mathbf z}),
\label{operator_psi}
\end{equation}
and
\begin{equation}
{\mathop{\rm \chi}\nolimits}[f]({\mathbf z})= \frac{{\mathop{\rm I}\nolimits}}{{\mathop{\rm I}\nolimits}-{\mathop{\rm L}\nolimits}}[f] ({\mathbf z}).
\label{operator_chi}
\end{equation}
Finally, it follows that the incident collision densities satisfies the stationary integral transport equation
\begin{equation}
\Psi({\mathbf z}|{\mathbf z}_0)={\mathop{\rm K}\nolimits}\Psi({\mathbf z}|{\mathbf z}_0)+{\mathop{\rm T}\nolimits}{\cal S}
\label{integral_transport_equations_one}
\end{equation}
whereas the outgoing collision density satisfies
\begin{equation}
\chi({\mathbf z}|{\mathbf z}_0)={\mathop{\rm L}\nolimits}\chi({\mathbf z}|{\mathbf z}_0)+{\cal S}.
\label{integral_transport_equations_two}
\end{equation}
The solutions $\Psi({\mathbf z}|{\mathbf z}_0)$ and $\chi({\mathbf z}|{\mathbf z}_0)$ are related by
\begin{equation}
\chi({\mathbf z}|{\mathbf z}_0)={\mathop{\rm C}\nolimits}\Psi({\mathbf z}|{\mathbf z}_0)+{\cal S}.
\label{link_chi_psi}
\end{equation}
Observe that $\chi({\mathbf z}|{\mathbf z}_0)$ obeys an integral equation whose source term is the physical source ${\cal S}$, whereas the source term in the equation for $\Psi({\mathbf z}|{\mathbf z}_0)$ is the so-called first-collision source ${\mathop{\rm T}\nolimits}{\cal S}$, i.e., the density of particles entering the first collision. For reasons that will be clear later, it is expedient to introduce the function $\varphi({\mathbf z}|{\mathbf z}_1)$, being the solution of $({\mathop{\rm I}\nolimits}-{\mathop{\rm K}\nolimits})\varphi({\mathbf z}|{\mathbf z}_1)={\cal S}'$, for a point-source consisting in a particle entering the first collision at ${\mathbf z}_1$. Then, the incident collision density can be expressed by the convolution
\begin{equation}
\Psi({\mathbf z}|{\mathbf z}_0)=\int \varphi({\mathbf z}|{\mathbf z}_1)T({\mathbf z}_1|{\mathbf z}_0)d{\mathbf z}_1.
\label{convolution}
\end{equation}

\begin{figure}[t]
   \centerline{ \epsfclipon \epsfxsize=9.0cm
\epsfbox{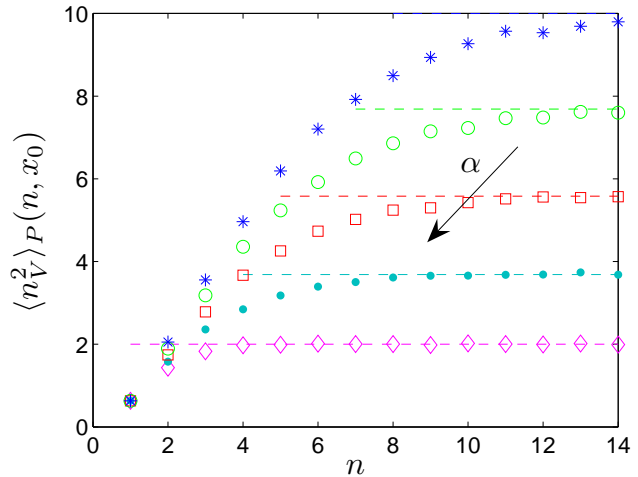} }
   \caption{(Color online) Second moment of the collision number $\langle n^2_V\rangle_{{P}}(n,x_0)$ for a sphere with radius $R=1$ and leakage boundary conditions. The source is isotropic ($x_0=0$), and $p=1$. Dashed lines: Eq.~\eqref{moment_two_aniso}. Symbols: Monte Carlo simulations. The persistence parameter is $\alpha=0$ (dark blue stars), $\alpha=0.25$ (green circles), $\alpha=0.5$ (red squares), $\alpha=0.75$ (light blue dots), and $\alpha=1$ (violet diamonds).}
   \label{fig4}
\end{figure}

\section{Collision statistics}
\label{direct_section}

Suppose that the trajectories of the random flights described above are observed until the walker either disappears by leaving an external boundary, or is absorbed.

\subsection{Scattering and absorption events}

We first assume that each event in a given volume $V$ is detected when the particle enters a collision (in other words, we do not discriminate scattering and absorption events). The quantity $q_{\Psi}(n|{\mathbf z}_0)=\int_V d{\mathbf z} \Psi({\mathbf z},n|{\mathbf z}_0)$ represents the survival probability, i.e., the probability for the particle to be in $V$ up to entering the $n$-th collision. From the Markovian nature of the process ${\mathbf z}$~\cite{zdm_pre}, it follows that the probability of detecting $n_V$ collision events in the volume $V$ is
\begin{equation}
{P}(n_{V}|{\mathbf z}_0)=q_{\Psi}(n_V|{\mathbf z}_0)-q_{\Psi}(n_V+1|{\mathbf z}_0).
\label{pnv}
\end{equation}
The moments are given by
\begin{equation}
\langle n_V^m \rangle_{{P}}({\mathbf z}_0) = \sum_{n_{V}=1}^{+\infty} n_{V}^m {P}(n_{V}|{\mathbf z}_0)
\label{n_poisson}
\end{equation}
for $m \ge 1$, and depend on the boundary conditions on $\partial V$, which affect the functional form of the propagator~\cite{zoia_dumonteil_mazzolo}. Setting boundary conditions at infinity corresponds to defining a fictitious (`transparent') volume $V$, where particles can indefinitely cross $\partial V$ back and forth. On the contrary, the use of leakage boundary conditions, i.e., when the particle is lost upon crossing the boundary, leads to the formulation of first-passage problems~\cite{redner, zdm_prl}.

Normalization implies $\langle n^0_V \rangle_{{P}}({\mathbf z}_0) = 1$, and direct calculation from Eqs.~\eqref{n_poisson} and~\eqref{pnv} yields
\begin{eqnarray}
\langle n^1_V \rangle_{{P}}({\mathbf z}_0) = \int_V d {\mathbf z} \Psi [{\cal S}]=\int_V d {\mathbf z}\Psi({\mathbf z}|{\mathbf z}_0).
\label{average_np}
\end{eqnarray}
Observe that the integral of the collision density over a volume $V$ gives the mean number of collisions within that domain, hence the name given to $\Psi({\mathbf z}|{\mathbf z}_0)$.

\begin{figure}[t]
   \centerline{ \epsfclipon \epsfxsize=9.0cm
\epsfbox{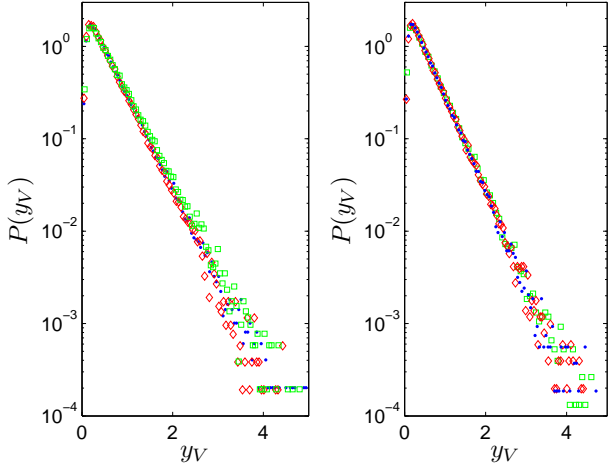} }
   \caption{(Color online) The distribution ${P}(y_{V})$ for $x_0=0$ as a function of the rescaled variable $y_V=n_V / (\eta R^2) $. Left: $R=30$ and $\alpha=0.25$ (blue dots), $\alpha=0.5$ (green squares), and $\alpha=0.75$ (red diamonds). Right: $\alpha=0.5$ and $R=30$ (blue dots), $R=40$ (green squares), and $R=50$ (red diamonds).}
   \label{fig5}
\end{figure}

Higher order moments of $n_V$ follow from Eqs.~\eqref{psi_n} and~\eqref{n_poisson}, and read
\begin{equation}
\langle n_V^m \rangle_{{P}}({\mathbf z}_0) =\sum_{n_{V}=1}^{+\infty} n_{V}^m \int_V d{\mathbf z} {\mathop{\rm K}\nolimits}^{n_V-1}({\mathop{\rm I}\nolimits}-{\mathop{\rm K}\nolimits}){\mathop{\rm T}\nolimits}[{\cal S}],
\label{moment_2}
\end{equation}
for $m \ge 1$. By resorting to the identity~\eqref{identity_one}, we obtain then
\begin{equation}
\langle n_V^m \rangle_{{P}}({\mathbf z}_0) = \sum_{k=1}^{m}k!S_{m,k} \int_V d{\mathbf z} {\mathop{\rm \Psi}\nolimits} \left( {\mathop{\rm C}\nolimits} {\mathop{\rm \Psi}\nolimits} \right)^{k-1}[{\cal S}],
\label{moments_operator}
\end{equation}
where $S_{m,k}$ are the Stirling numbers of second kind (see Eq.~\eqref{stirling_two}). We can further introduce the factorial moments $\langle n_V^{(m)} \rangle({\mathbf z}_0)$, where $x^{(k)}=x(x+1)...(x+k-1)$ is the rising factorial~\cite{erdelyi}. The factorial moments are related to the moments by
\begin{equation}
\langle n_V^{(m)} \rangle({\mathbf z}_0) = \sum_{k=0}^{m}|s_{m,k}|\langle n_V^{k} \rangle({\mathbf z}_0),
\end{equation}
$s_{m,k}$ being the Stirling numbers of first kind (see Eq.~\eqref{stirling_one}). From the operator identity~\eqref{identity_two}, we have then
\begin{equation}
\langle n_V^{(m)} \rangle_{{P}}({\mathbf z}_0) = m! \int_V d{\mathbf z} \left( \frac{{\mathop{\rm I}\nolimits}}{{\mathop{\rm I}\nolimits}-{\mathop{\rm K}\nolimits}}\right)^m {\mathop{\rm T}\nolimits}[{\cal S}],
\label{moment_2_factorial}
\end{equation}
which holds for $m \ge 1$. Setting now
\begin{equation}
\langle n_V^{(m)} \rangle'_{{P}}({\mathbf z}_1) = m! \int_V d{\mathbf z} \left( \frac{{\mathop{\rm I}\nolimits}}{{\mathop{\rm I}\nolimits}-{\mathop{\rm K}\nolimits}}\right)^m [{\cal S}']
\label{moment_prime_factorial}
\end{equation}
yields the recursion property for the factorial moments
\begin{equation}
\langle n_V^{(m)} \rangle'_{{P}}({\mathbf z}_1) = m \int_V d{\mathbf z} \varphi({\mathbf z}|{\mathbf z}_1) \langle n_V^{(m-1)} \rangle'_{{P}}({\mathbf z}),
\label{factorial_recursion_psi}
\end{equation}
for $m \ge 1$, starting from $\langle n_V^{(0)} \rangle'_{{P}}({\mathbf z}_1)=1$. Then, from Eq.~\eqref{convolution} we have
\begin{equation}
\langle n_V^{(m)} \rangle_{{P}}({\mathbf z}_0) = \int \langle n_V^{(m)} \rangle'_{{P}}({\mathbf z}_1) T({\mathbf z}_1|{\mathbf z}_0)d{\mathbf z}_1.
\label{mom_prime_mom}
\end{equation}

\subsection{Scattering events}

Suppose now that events in $V$ are detected at the exit of each collision (in other words, only scattering collisions and the source ${\cal S}$ are recorded). The quantity $q_{\chi}(n|{\mathbf z}_0)=\int_V d{\mathbf z} \chi({\mathbf z},n|{\mathbf z}_0)$, $n \ge 0$, represents the probability for the particle to be in $V$ after having undergone the $n$-th collision. From the same argument as above, the probability of detecting $n_V$ scattering events in the volume $V$ is
\begin{equation}
{Q}(n_{V}|{\mathbf z}_0)=q_{\chi}(n_V-1|{\mathbf z}_0)-q_{\chi}(n_V|{\mathbf z}_0).
\label{qnv}
\end{equation}
The moments are then
\begin{equation}
\langle n_V^m \rangle_{{Q}}({\mathbf z}_0) = \sum_{n_{V}=1}^{+\infty} n_{V}^m {Q}(n_{V}|{\mathbf z}_0)
\label{n_q}
\end{equation}
for $m \ge 1$, with $\langle n^0_V \rangle_{{Q}}({\mathbf z}_0) = 1$ from normalization. Direct calculation from Eqs.~\eqref{n_q} and~\eqref{qnv} yields
\begin{eqnarray}
\langle n^1_V \rangle_{{Q}}({\mathbf z}_0) = \int_V d {\mathbf z} \chi [{\cal S}]= \int_V d {\mathbf z} \chi({\mathbf z}|{\mathbf z}_0).
\label{average_nq}
\end{eqnarray}

Higher order moments of $n_V$ follow from Eq.~\eqref{chi_n}, and read
\begin{equation}
\langle n_V^m \rangle_{{Q}}({\mathbf z}_0) =\sum_{n_{V}=1}^{+\infty} n_{V}^m \int_V d{\mathbf z} {\mathop{\rm L}\nolimits}^{n_V-1}({\mathop{\rm I}\nolimits}-{\mathop{\rm L}\nolimits})[{\cal S}],
\label{moment_2_chi}
\end{equation}
which holds for $m \ge 1$. By resorting to the identity~\eqref{identity_one}, we obtain
\begin{equation}
\langle n_V^m \rangle_{{Q}}({\mathbf z}_0) = \sum_{k=1}^{m}k!S_{m,k} \int_V d{\mathbf z} {\mathop{\rm \chi}\nolimits} \left( {\mathop{\rm L}\nolimits} {\mathop{\rm \chi}\nolimits} \right)^{k-1}[{\cal S}].
\label{moments_chi}
\end{equation}
As done before, we can further introduce the factorial moments $\langle n_V^{(m)} \rangle({\mathbf z}_0)_{{Q}}$. From the  identity~\eqref{identity_two}, for $m \ge 1$ we have then
\begin{equation}
\langle n_V^{(m)} \rangle_{{Q}}({\mathbf z}_0) = m! \int_V d{\mathbf z} \left( \frac{{\mathop{\rm I}\nolimits}}{{\mathop{\rm I}\nolimits}-{\mathop{\rm L}\nolimits}}\right)^m [{\cal S}],
\label{moment_2_factorial_chi}
\end{equation}
which finally yields the recursion property for the factorial moments
\begin{equation}
\langle n_V^{(m)} \rangle_{{Q}}({\mathbf z}_0) = m \int_V d{\mathbf z} \chi({\mathbf z}|{\mathbf z}_0) \langle n_V^{(m-1)} \rangle_{{Q}}({\mathbf z}),
\label{factorial_recursion_chi}
\end{equation}
for $m \ge 1$. Similar results for the factorial moments appear in~\cite{pitman} (and references therein) under the name of discrete Feynman moment formula.

\section{Diffusion limit}
\label{diff_lim}

Suppose that the walker evolves in a medium without boundaries, starting from an isotropic source. Assume for the sake of simplicity that there is no absorption, and the speed $v=|{\mathbf v}|$ is constant. The spread at the $n$-th collision $\langle r^2 \rangle(n)=\int d{\mathbf r}\int d{\mathbf v} |{\mathbf r}-{\mathbf r}_0|^2 \Psi({\mathbf r},{\mathbf v},n|{\mathbf r}_0,{\mathbf v}_0)$ reads
\begin{equation}
\langle r^2 \rangle(n)=n\langle \ell^2 \rangle + 2n\langle \ell \rangle^2 \frac{\mu}{1-\mu} - 2\langle \ell \rangle^2 \frac{\mu\left( 1-\mu^n\right) }{\left( 1-\mu\right) ^2},
\label{spread_n}
\end{equation}
where $\ell$ is the inter-collision length, and $0<\mu<1$ is the average (polar) scattering angle~\cite{gandj}. Observe that Eq.~\eqref{spread_n} does not explicitly depend on the dimension $d$ of the embedding space. When the typical displacements are much smaller than the domain size, say $\langle \ell^2 \rangle\ll R^2$, $n$ is large, and the last term in Eq.~\eqref{spread_n} can be dropped. If the two first moments of the flight length distribution are finite, we can set $\gamma= \langle \ell \rangle^2 / \langle \ell^2 \rangle > 0$, and rewrite
\begin{equation}
\langle r^2 \rangle(n) \simeq n\langle \ell^2 \rangle \left(1+ 2\gamma \frac{\mu}{1-\mu}\right).
\label{spread_n_regular}
\end{equation}
Equation~\eqref{spread_n_regular} implies that at large $n$ any anisotropic walk will behave like an isotropic walk (with a linear spread), provided that we rescale $n$ by a factor $\eta=1/\left[ 1+ 2\gamma \mu/(1-\mu)\right] $ that depends on the specific features of the jump length distribution. For a sharply forward-peaked walk ($\mu \simeq 1$), the number of collisions needed for attaining isotropy becomes very large as compared to an isotropic walk, for a given $\langle \ell^2 \rangle$. When $n$ is sufficiently large, we expect the same scaling to carry over to the moments, i.e.,
\begin{equation}
\langle n_V^{m} \rangle_P({\mathbf z}_0)_{aniso} \simeq \eta^m\langle n_V^{m} \rangle_P({\mathbf z}_0)_{iso}.
\label{eta_aniso_scaling}
\end{equation}

The so-called diffusion limit is reached when $\langle \ell^2 \rangle\ll R^2$ and the average flight time $\tau=\langle \ell \rangle / v$ is vanishing small, while preserving a constant ratio $D=\langle \ell^2 \rangle / \tau$, namely the diffusion coefficient. Under these conditions, the collision number in $V$ diverges, whereas the quantity $t_{V} = \sum_{i=1}^{n_{V}} \ell_i/v$ converges to the residence time in the volume. Actually, $t_{V}$ should take into account also additional terms due to boundary conditions. However, as $\tau \to 0$, the trajectory will almost surely have a turning point touching the boundary, so that corrections can be safely neglected. By following the arguments in~\cite{zdm_prl}, the moments $\langle t^m_{V} \rangle ({\mathbf r}_0)$ of the residence time will be given by the celebrated Kac formula for the isotropic Brownian motion~\cite{kac, berezhkovskii}, up to a scaling factor $\eta^m$.

\section{The rod model}
\label{rod_sec}

The approach presented in Sec.~\ref{direct_section} allows explicitly evaluating the moments $\langle n_V^m \rangle({\mathbf r}_0,{\mathbf v}_0)$. When the equilibrium distribution is known, this amounts to solving the convolution integrals in Eqs.~\eqref{moments_operator} and~\eqref{moments_chi}. However, analytical expressions for $\Psi({\mathbf r},{\mathbf v}|{\mathbf r}_0,{\mathbf v}_0)$ can be obtained only in a few cases~\cite{zoia_dumonteil_mazzolo, zdm_prl, orsingher, kolesnik}, so that one must generally resort to numerical integration. In this Section we exemplify the moments formulas above for a well-known and long-studied system where calculations can be carried out analytically.

\subsection{Exponential flights}
\label{exponential_flights}

When the scattering centers are spatially uniform, i.e., when the traversed medium is homogeneous, the inter-collision lengths are exponentially distributed. Exponential flights describe for instance the displacements of neutral particles (neutrons, photons) in matter, the motion of electrons in semiconductors, the migration of biological species (often called velocity jump process), and are widely used in gas dynamics (Lorentz gas)~\cite{blanco_fournier, mazzolo, velocity_jump, bacteria, weiss_review, wing, lorentz}. The displacement kernel for exponential flights reads
\begin{equation}
T({\mathbf r}|{\mathbf r'},{\mathbf v})=\Sigma_t({\mathbf r'},{\mathbf v})e^{-\int_{0}^{{\mathbf \omega}\cdot ({\mathbf r}-{\mathbf r'})}\Sigma_t({\mathbf r'}+s{\mathbf \omega},{\mathbf v})ds},
\label{exp_kernel}
\end{equation}
when ${\mathbf \omega}={\mathbf v}/v$ is parallel to ${\mathbf r}-{\mathbf r'}$, oriented as ${\mathbf v}$~\cite{spanier, lux}. The quantity $\Sigma_t({\mathbf r},{\mathbf v})$ is the total cross section, which is proportional to the probability of particle-medium interaction along a straight line (it carries units of the inverse of a length). Combining Eqs.~\eqref{exp_kernel} and~\eqref{integral_transport_equations_one} yields
\begin{eqnarray}
{\mathbf \omega}\cdot \nabla \phi({\mathbf r},{\mathbf v}|{\mathbf r}_0,{\mathbf v}_0)+\Sigma_t({\mathbf r},{\mathbf v})\phi({\mathbf r},{\mathbf v}|{\mathbf r}_0,{\mathbf v}_0)=\nonumber \\
=\int d{\mathbf v'}C({\mathbf v}|{\mathbf v'},{\mathbf r}) \Sigma_t({\mathbf r},{\mathbf v'})\phi({\mathbf r},{\mathbf v'}|{\mathbf r}_0,{\mathbf v}_0)+{\cal S}
\label{phi_exp_kernel}
\end{eqnarray}
where $\Psi({\mathbf r},{\mathbf v}|{\mathbf r}_0,{\mathbf v}_0) = \Sigma_t({\mathbf r},{\mathbf v})\phi({\mathbf r},{\mathbf v}|{\mathbf r}_0,{\mathbf v}_0)$~\cite{spanier, lux}. The quantity $\phi({\mathbf r},{\mathbf v}|{\mathbf r}_0,{\mathbf v}_0)$ is called flux in neutronics.

In the following, we introduce some simplifying hypotheses. First, we consider a $1d$ setup, where particles undergo exponential displacements along a straight line, only forward and backward directions being allowed: the so-called {\em rod model}~\cite{wing, weiss, hughes}. Further, we assume that the particle energy is preserved along each trajectory, which corresponds to the so-called one-speed approximation. Finally, we take the total cross section and the scattering probability to be constant, i.e., $\Sigma_t({\mathbf r},{\mathbf v})=\Sigma_t$ and $p({\mathbf r},{\mathbf v})=p$. Without loss of generality, we set $\Sigma_t=1$. Despite being admittedly oversimplified, this model can nonetheless capture the essential features of the corresponding physical system.

We define $\omega_f$ and $\omega_b$ the forward and backward directions, respectively. Similarly, we denote by ${\cal S}_f$ and ${\cal S}_b$ the forward and backward intensities of the source, located at $x_0$. Anisotropy is taken into account by introducing a persistence coefficient $\alpha$ such that after each scattering collision the particle preserves its direction $\omega$ with probability $\alpha$, and inverses its direction otherwise~\cite{weiss, weiss_tele}. This imposes
\begin{equation}
C(\omega|\omega')=p \left[ \alpha \delta(\omega-\omega')+(1-\alpha)\delta(\omega+\omega') \right] .
\end{equation}
The persistence coefficient is related to the average scattering angle by $-1+2\alpha=\mu$. Remark that $\alpha=1$ corresponds to a walker that systematically preserves its incident direction (forward scattering), whereas $\alpha=0$ to a walker that systematically reverses its incident direction. The model with $\alpha=0$ has long been investigated under the name of telegrapher's equation~\cite{weiss, kac_telegraph, weiss_tele}. Isotropic scattering is recovered for $\alpha=1/2$ ($\mu=0$). For exponential flights $\gamma=1/2$, so that we have $\eta=1-\mu$. The volume $V$ is assumed to be the interval $V=[-R,R]$. With this choice of parameters and notations, Eq.~\eqref{phi_exp_kernel} yields the following set of stationary first-order differential equations for the incident collision density
\begin{widetext}
\begin{eqnarray}
\left( \frac{\partial}{\partial x} +1\right) \Psi(x,\omega_f|x_0,\omega_f)=p\left[\alpha \Psi(x,\omega_f|x_0,\omega_f)+(1-\alpha)\Psi(x,\omega_b|x_0,\omega_f)\right]  + {\cal S}_f,\nonumber \\
\left( -\frac{\partial}{\partial x} +1\right) \Psi(x,\omega_b|x_0,\omega_f)=p\left[\Psi(x,\omega_b|x_0,\omega_f)+(1-\alpha)\Psi(x,\omega_f|x_0,\omega_f) \right],\nonumber \\
\left( \frac{\partial}{\partial x} +1\right) \Psi(x,\omega_f|x_0,\omega_b)=p\left[\alpha\Psi(x,\omega_f|x_0,\omega_b)+(1-\alpha)\Psi(x,\omega_b|x_0,\omega_b) \right], \nonumber \\
\left( -\frac{\partial}{\partial x} +1\right) \Psi(x,\omega_b|x_0,\omega_b)=p\left[\alpha\Psi(x,\omega_b|x_0,\omega_b)+(1-\alpha)\Psi(x,\omega_f|x_0,\omega_b) \right] + {\cal S}_b.
\label{eq_rod_f_b}
\end{eqnarray}
\end{widetext}
These equations are linear and can be put in matricial form: together with appropriate boundary conditions, this leads to an explicit solution for $\Psi(x,\omega|x_0,\omega_0)$. However, the solution may not exist, depending on the choice of the equation parameters. Actually, $1d$ exponential flights are recurrent walks (i.e., they almost surely re-visit their initial position~\cite{zoia_dumonteil_mazzolo, zdm_prl}), so that $\Psi(x,\omega|x_0,\omega_0)$ diverges when $p=1$, unless leakage boundary conditions are imposed, i.e., the particles are lost upon crossing the boundary of the domain.

\subsection{Examples of calculations}
\label{examples}

The effects of the scattering probability $p$ on the moments have been discussed elsewhere~\cite{zdm_pre}. Here we will focus on the case of leakage boundary conditions without absorption, which allows emphasizing the effects of anisotropy. Leakages at $x=\pm R$ impose $\Psi(-R,\omega_f|x_0,\omega_f)=0$, $\Psi(R,\omega_b|x_0,\omega_f)=0$, $\Psi(R,\omega_b|x_0,\omega_b)=0$, and $\Psi(-R,\omega_f|x_0,\omega_b)=0$, which corresponds to an homogeneous medium surrounded by vacuum. We choose to observe events when particles enter collisions, which amounts to counting scattering and absorptions, and disregarding the contribution of the source. Once $\Psi(x,\omega|x_0,\omega_0)$ is known, the moments $\langle n_V^{m} \rangle_{{P}}(x_0)$ can be obtained from Eq.~\eqref{moments_operator}. For instance, the average $n_V$ reads
\begin{equation}
\langle n_V^{1} \rangle_{{P}}(x_0) = \frac{2R +\eta R^2 -\eta x_0^2}{2} ,
\label{moment_one_aniso}
\end{equation}
when the source is isotropic. This expression explicitly depends on the initial position $x_0$, on the persistence coefficient $\alpha$ and on the size $R$ of the domain, when $\Sigma_t=1$. Moments for generic $\Sigma_t$ are simply obtained by rescaling the space variables $R$ and $x_0$ by a factor $\Sigma_t$. In Fig.~\ref{fig1} we display the behavior of $\langle n_V^{1} \rangle_P(x_0)$ for $R=1$. Remark that the average has always a maximum at $x_0=0$ (whose height decreases with $\alpha$) and a minimum at $x_0=R$ (independent of $\alpha$). When $\alpha=1$, the average collision number becomes independent of the starting point $x_0$ (recall that we have chosen an isotropic source). The moment $\langle n_V^{1} \rangle_{{Q}}(x_0)$ can be computed based on Eqs.~\eqref{average_nq} and~\eqref{link_chi_psi}, and reads
\begin{equation}
\langle n_V^{1} \rangle_{{Q}}(x_0) = \langle n_V^{1} \rangle_{{P}}(x_0) + 1.
\end{equation}
When $R$ is large (i.e., when $R \Sigma_t \gg 1 $), by comparing Eq.~\eqref{moment_one_aniso} with the results for isotropic transport in~\cite{zdm_prl}, we have $\langle n_V^{1} \rangle_P(x_0)_{aniso} \simeq \eta \langle n_V^{1} \rangle_P(x_0)_{iso}$, in agreement with Eq.~\eqref{eta_aniso_scaling}.

The second moment can be also easily obtained by direct integration, and the formula reads
\begin{widetext}
\begin{eqnarray}
\langle n_V^{2} \rangle_{{P}}(x_0)= \langle n_V^{1} \rangle_P(x_0) + \frac{24R^2 + 20 \eta R^3 + 5 \eta^2 R^4 -
          12 \eta R x_0^2  -6\eta^2 R^2x_0^2 + \eta^2 x_0^4}{12} +(1-\eta)\left(x_0^2-R^2 \right) .  
\label{moment_two_aniso}
\end{eqnarray}
\end{widetext}
In Fig.~\ref{fig2} we display the behavior of $\langle n_V^{2} \rangle_{{P}}(x_0)$ for $R=1$ and an isotropic source. In this case, the curves are not monotonic with respect to $x_0$: by varying $\alpha$, the maximum of $\langle n_V^{2} \rangle_{{P}}(x_0)$ is either at the center of the domain, or at its boundary. The moment $\langle n_V^{2} \rangle_{{Q}}(x_0)$ can be computed based on Eqs.~\eqref{moments_chi} and~\eqref{link_chi_psi}, and reads
\begin{equation}
\langle n_V^{2} \rangle_{{Q}}(x_0) = \langle n_V^{2} \rangle_{{P}}(x_0) + 2\langle n_V^{1} \rangle_{{P}}(x_0) +1.
\end{equation}
Again, by comparison with the results for isotropic transport, when $R$ is large we have $\langle n_V^{2} \rangle_P(x_0)_{aniso} \simeq \eta^2 \langle n_V^{2} \rangle_P(x_0)_{iso}$, in agreement with Eq.~\eqref{eta_aniso_scaling}.

Then, for the same geometrical configuration (i.e., leakage boundaries and $R=1$), we analyze the behavior of the moments when observed up to entering the $n$-th collision, as a function of $n$ for $x_0=0$ and varying $\alpha$. These quantities, that we denote by $\langle n_V^{m} \rangle_{{P}}(n,x_0)$, are easily obtained by Monte Carlo simulation, and are expected to converge to the respective $\langle n_V^{m} \rangle_{{P}}(x_0)$ when $n \to \infty$. Figures~\ref{fig3} and~\ref{fig4} show that the average and the second moment grow with $n$ and saturate to their respective asymptotic values, given by Eqs.~\eqref{moment_one_aniso} and~\eqref{moment_two_aniso}. The number of collisions needed to reach saturation, as well as the asymptotic value at saturation, decrease with increasing $\alpha$. This can be understood by considering that a forward-peaked walker ($\alpha \simeq 1$) will undergo fewer collisions in $V$ (before crossing the boundary) than a walker with small $\alpha$, which on the contrary frequently reverses its direction and thus stays longer in $V$.

Finally, we conclude with some considerations concerning the limit behavior of the collision number distribution. To fix the ideas, let us assume that $x_0=0$. When $R$ is large, by direct inspection we recognize the scaling $\langle n_V^{m} \rangle_{{P}} \simeq c_m \eta^m R^{2m}$, where $c_m=(-1)^m m! {\mathop{\rm E}\nolimits}_{2m} / (2m)!$, and
\begin{equation}
{\mathop{\rm E}\nolimits}_{2n}=i \sum^{2n+1}_{k=1} \sum^{k}_{j=0} (-1)^{j}\binom {k} {j} \frac{\left(k-2j \right)^{2n+1} }{k(2i)^k }
\end{equation}
are the Euler's numbers~\cite{erdelyi}. The moment generating function $G(u)$ associated to ${P}(n_{V})$ has the moment expansion
\begin{equation}
G(u)=\sum^{+\infty}_{m=0} \langle n_V^{m} \rangle_{{P}} \frac{(-u)^m}{m!}=\sum^{+\infty}_{m=0} \frac{{\mathop{\rm E}\nolimits}_{2m}}{(2m)!}\left( R^2\eta u\right) ^m.
\end{equation}
By carrying out the sum we get $G(u)={\mathop{\rm sech}\nolimits}(R\sqrt{\eta u})$, where we recognize the scaling variable $y_V=n_V / (\eta R^2) $. In Fig.~\ref{fig5} we display ${P}(y_{V})$ as a function of $y_{V}$, for various values of $\eta$ and $R$: it is immediately apparent that all the curves collapse. The exponential tail of ${P}(y_{V})$ for large $y_V$ is expected on the basis of the Tauberian theorems~\cite{feller}, since $G(u) \simeq 1 - \eta R^2 u/2$ for small $u$.

All analytical results reported here have been validated by comparison with Monte Carlo simulations with $10^6$ particles.

\section{Conclusions}
\label{conclusions}

In this paper we have proposed a general approach to the counting statistics for the number of events falling within a given region $V$ of the phase space, when the underlying process is a random flight. By resorting to survival probabilities, we have provided an explicit description of the moments and factorial moments of the collision number $n_V$. Only a minimal number of hypotheses on the underlying transport kernels $K({\mathbf z}|{\mathbf z'})$ or $L({\mathbf z}|{\mathbf z'})$ are required, and we were able to take into account the effects of jump length distribution, anisotropy, absorption and boundary conditions. In this work we have in particular focused on anisotropy, and discussed some examples of analytical calculations for the class of random flights where displacements are exponentially distributed, namely, the exponential flights.

In view of the physical systems to which random flights most often apply, e.g., gas dynamics and neutronics, we have found natural here to resort to the language specific to stochastic transport phenomena. However, this same formalism could be further generalized in terms of semi-Markov renewal processes for an arbitrary state variable ${\mathbf q}$ evolving in the phase space according to some transition kernel.

Finally, observe that we might take advantage of the knowledge on the number of collisions in a given domain, when available, as an estimator to infer the equilibrium distribution of the underlying stochastic path, which is seldom directly accessible. Indeed, while for instance in neutronics the underlying transport process is supposedly known, and one is typically interested in assessing the collision statistics (e.g., the deposited energy and/or the radiation damage), in life sciences one could measure the number of hits of the displacing species in a region $V$ so as to probe its possibly unknown dynamics. Nonetheless, this inverse problem might be ill-posed, and deserves further investigation.

\appendix

\section{Operator identities}

In this Section we recall some identities that are used in the paper. The formal Neumann series is defined by
\begin{equation}
\sum_{n=0}^{\infty} {\mathop{\rm A}\nolimits}^{n}[f]({\mathbf z})= \frac{{\mathop{\rm I}\nolimits}}{{\mathop{\rm I}\nolimits}-{\mathop{\rm A}\nolimits}}[f]({\mathbf z}),
\label{neumann_def}
\end{equation}
where ${\mathop{\rm I}\nolimits}$ is the identity operator.

For any sufficiently well-behaved operator ${\mathop{\rm A}\nolimits}$, we have the identity
\begin{eqnarray}
({\mathop{\rm I}\nolimits}-{\mathop{\rm A}\nolimits})\sum_{n=0}^{+\infty} n^m {\mathop{\rm A}\nolimits}^{n-1}[f]({\mathbf z})=\nonumber \\
=\frac{{\mathop{\rm I}\nolimits}}{{\mathop{\rm I}\nolimits}-{\mathop{\rm A}\nolimits}}\sum_{k=1}^{m}k!S_{m,k} \left( \frac{{\mathop{\rm A}\nolimits}}{{\mathop{\rm I}\nolimits}-{\mathop{\rm A}\nolimits}}\right)^{k-1} [f]({\mathbf z}),
\label{identity_one}
\end{eqnarray}
where the coefficients
\begin{equation}
S_{m,k}=\frac{1}{k!} \sum_{i=0}^{k} (-1)^i \binom {k} {i} \left( k-i\right)^m
\label{stirling_two}
\end{equation}
are the Stirling numbers of second kind~\cite{erdelyi}. Moreover, we have
\begin{eqnarray}
\sum_{k=0}^{m} |s_{m,k}| \sum_{n=0}^{+\infty} n^k {\mathop{\rm A}\nolimits}^{n-1} ({\mathop{\rm I}\nolimits}-{\mathop{\rm A}\nolimits}) [f]({\mathbf z})=\nonumber \\
=m!\left( \frac{{\mathop{\rm I}\nolimits}}{{\mathop{\rm I}\nolimits}-{\mathop{\rm A}\nolimits}}\right)^m [f]({\mathbf z}),
\label{identity_two}
\end{eqnarray}
where the coefficients $s_{m,k}$ are the Stirling numbers of the first kind~\cite{erdelyi}, which are defined as the coefficients in the expansion
\begin{equation}
(x)_m =\sum_{k=0}^{m} s_{m,k} x^k,
\label{stirling_one}
\end{equation}
$(x)_m$ being the falling factorial
\begin{equation}
(x)_m= x(x-1)(x-2)...(x-m+1).
\end{equation}

\acknowledgments

The authors wish to thank Dr.~F.~Malvagi for useful discussions.

\end{document}